\documentstyle[12pt]{article}

\newcommand{\sect}[1]{\setcounter{equation}{0}\section{#1}}

\def\bseq{\begin{subequation}}  
\def\eseq{\end{subequation}}
\def\bsea{\begin{subeqnarray}}  
\def\esea{\end{subeqnarray}}

\def\Bar#1{\overline{#1}}                       

\newcommand{\beq}{\begin{equation}}
\newcommand{\eeq}{\end{equation}}
\newcommand{\bea}{\begin{eqnarray}}
\newcommand{\eea}{\end{eqnarray}}
\newcommand{\ena}{\end{eqnarray}}

\newcommand {\non}{\nonumber}

\renewcommand{\a}{\alpha}
\renewcommand{\b}{\beta}

\renewcommand{\d}{\delta}
\newcommand{\th}{\theta}

\newcommand{\pa}{\partial}

\newcommand{\g}{\gamma}

\newcommand{\m}{\mu}

\newcommand{\n}{\nu}

\newcommand{\s}{\sigma}
\renewcommand{\S}{\Sigma}

\renewcommand{\o}{\omega}

\def\Mb{\kern 2pt\mathchoice
            {
             \vbox{\hrule width10pt height 0.4pt depth 0pt
                 \kern 1.2pt\hbox{\kern -2pt$\displaystyle M$}}}
            {
                 \vbox{\hrule width10pt height 0.4pt depth 0pt
                 \kern 1.2pt\hbox{\kern -2pt$\textstyle M$}}}
            {
\vbox{\hrule width6pt height 0.4pt depth 0pt
                 \kern 1.0pt\hbox{\kern -2pt$\scriptstyle M$}}}
            {
                 \vbox{\hrule width5pt height 0.4pt depth 0pt
                 \kern 0.8pt\hbox{\kern -2pt$\scriptscriptstyle M$}}}}

\def\Sb{\kern 2pt\mathchoice
            {
                 \vbox{\hrule width6pt height 0.4pt depth 0pt
                 \kern 1.2pt\hbox{\kern -2pt$\displaystyle S$}}}
            {
                 \vbox{\hrule width6pt height 0.4pt depth 0pt
                 \kern 1.2pt\hbox{\kern -2pt$\textstyle S$}}}
            {
                 \vbox{\hrule width3.5pt height 0.4pt depth 0pt
                 \kern 1.0pt\hbox{\kern -2pt$\scriptstyle S$}}}
            {
                 \vbox{\hrule width3pt height 0.4pt depth 0pt
                 \kern 0.8pt\hbox{\kern -2pt$\scriptscriptstyle S$}}}}

\def\Rb{\kern 2pt\mathchoice
            {
                 \vbox{\hrule width5.5pt height 0.4pt depth 0pt
                 \kern 1.2pt\hbox{\kern -2.5pt$\displaystyle R$}}}
            {
                 \vbox{\hrule width5.5pt height 0.4pt depth 0pt
                 \kern 1.2pt\hbox{\kern -2.5pt$\textstyle R$}}}
            {
                 \vbox{\hrule width3.5pt height 0.4pt depth 0pt
                 \kern 1.0pt\hbox{\kern -2.2pt$\scriptstyle R$}}}
            {
                 \vbox{\hrule width3pt height 0.4pt depth 0pt
                 \kern 0.8pt\hbox{\kern -2.2pt$\scriptscriptstyle R$}}}}

  \def\pp{{\mathchoice
          {
              \kern 1pt%
              \raise 1pt
              \vbox{\hrule width5pt height0.4pt depth0pt
                    \kern -2pt
                    \hbox{\kern 2.3pt
                          \vrule width0.4pt height6pt depth0pt
                          }
                    \kern -2pt
                    \hrule width5pt height0.4pt depth0pt}%
                    \kern 1pt
           }
            {
              \kern 1pt%
              \raise 1pt
              \vbox{\hrule width4.3pt height0.4pt depth0pt
                    \kern -1.8pt
                    \hbox{\kern 1.95pt
                          \vrule width0.4pt height5.4pt depth0pt
                          }
                    \kern -1.8pt
                    \hrule width4.3pt height0.4pt depth0pt}%
                    \kern 1pt
            }
            {
              \kern 0.5pt%
              \raise 1pt
              \vbox{\hrule width4.0pt height0.3pt depth0pt
                    \kern -1.9pt  
                    \hbox{\kern 1.85pt
                          \vrule width0.3pt height5.7pt depth0pt
                          }
                    \kern -1.9pt
                    \hrule width4.0pt height0.3pt depth0pt}%
                    \kern 0.5pt
            }
            {
              \kern 0.5pt%
              \raise 1pt
              \vbox{\hrule width3.6pt height0.3pt depth0pt
                    \kern -1.5pt
                    \hbox{\kern 1.65pt
                          \vrule width0.3pt height4.5pt depth0pt
                          }
                    \kern -1.5pt
                    \hrule width3.6pt height0.3pt depth0pt}%
                    \kern 0.5pt
            }
        }}

  \def\mm{{\mathchoice
                       {
                             \kern 1pt
               \raise 1pt    \vbox{\hrule width5pt height0.4pt depth0pt
                                  \kern 2pt
                                  \hrule width5pt height0.4pt depth0pt}
                             \kern 1pt}
                       {
                            \kern 1pt
               \raise 1pt \vbox{\hrule width4.3pt height0.4pt depth0pt
                                  \kern 1.8pt
                                  \hrule width4.3pt height0.4pt depth0pt}
                             \kern 1pt}
                       {
                            \kern 0.5pt
               \raise 1pt
                            \vbox{\hrule width4.0pt height0.3pt depth0pt
                                  \kern 1.9pt
                                  \hrule width4.0pt height0.3pt depth0pt}
                            \kern 1pt}
                       {
                           \kern 0.5pt
             \raise 1pt  \vbox{\hrule width3.6pt height0.3pt depth0pt
                                  \kern 1.5pt
                                  \hrule width3.6pt height0.3pt depth0pt}
                           \kern 0.5pt}
                       }}

\def\pd{{\kern0.5pt
                   + \kern-5.05pt \raise5.8pt\hbox{$\textstyle.$}\kern
0.5pt}}

\def\pmd{{\kern0.5pt
                  \pm \kern-5.05pt \raise6.3pt\hbox{$\textstyle.$}\kern1.5pt}}

\def\md{{\mathchoice
   {
      {{\kern 1pt - \kern-6.2pt \raise5pt\hbox{$\textstyle.$}\kern 1pt}}}
    {
      {{\kern 1pt - \kern-6.2pt \raise5pt\hbox{$\textstyle.$}\kern 1pt}}}
    {
      {\kern0.5pt - \kern-5.05pt \raise3.4pt\hbox{$\textstyle.$}\kern0.5pt}}
    {
      {\kern0.5pt - \kern-5.05pt \raise3.4pt\hbox{$\textstyle.$}\kern0.5pt}}}}

\newcommand{\ad}{{\dot{\alpha}}}
\newcommand{\bd}{{\dot{\beta}}}
\newcommand{\gd}{{\dot{\gamma}}}
\newcommand{\Del}{\nabla}
\newcommand{\Delb}{\bar{\nabla}}

\def\Sc{\scriptstyle}

\newcommand{\reff}[1]{(\ref{#1})}

\newcommand{\shalf}{{\Sc\frac{1}{2}}}
\newcommand{\sihalf}{{\Sc\frac{i}{2}}}

\newcommand{\squart}{{\Sc\frac{1}{4}}}

\renewcommand{\thefootnote}{\fnsymbol{footnote}}

\def\Bar#1{\overline{#1}}                       
\def\de{\nabla}

\begin{document}

\newpage
\begin{titlepage}
\begin{flushright}
{hep-th/9711120}\\
{BRX-TH-422}\\
{ITP-SB-97-56}\\
{WATPHYS-TH-97-10}
\end{flushright}
\vspace{2cm}
\begin{center}

{\large  A SUPERSPACE NORMAL COORDINATE DERIVATION
 OF THE DENSITY FORMULA}

Marcus T. ~Grisaru\footnote{grisaru@binah.cc.brandeis.edu. Supported in part by
NSF
Grant  PHY-960457}\\[.04in]
{\it  Physics Department,
Brandeis University\\
Waltham, MA 02254 USA}\\
[.08in]
Marcia E.~Knutt-Wehlau\footnote{knutt@physics.mcgill.ca.
Supported in part by a John Charles Polanyi Prize and a NSERC Postdoctoral
Fellowship.  Current Address: Physics Dept., McGill University, Montreal,
PQ  CANADA H3A 2T8}\\[.04in]
{\it Physics Department, University of Waterloo\\
Waterloo, ONT CANADA N2L 3G1}\\
{\rm {and}}\\[.08in]
Warren Siegel\footnote{siegel@insti.physics.sunysb.edu.
Supported in part by NSF Grant
PHY-9722101
}\\[.04in]
{\it Institute for Theoretical Physics, State University of New York\\
Stony Brook, NY 11794-3840 USA}

\vspace{1.0in}

{ABSTRACT}\\[.1in]
\end{center}
\begin{quote}
Using normal coordinate expansions we derive by purely superspace methods
the density formula giving  the  component action corresponding to a superspace
supergravity-matter action.
\end{quote}

\vfill

\begin{flushleft}
Nov. 1997
\end{flushleft}
\end{titlepage}

\newpage

\renewcommand{\thefootnote}{\arabic{footnote}}
\setcounter{footnote}{0}
\newpage
\pagenumbering{arabic}

\sect{Introduction}

In recent years, normal coordinate expansions in superspace have been used
for a variety of purposes. McArthur \cite{mcarthur}, extending to superspace
Spivak's
ordinary space discussion \cite{spivak}, has given a general expansion of the
vielbein and connection  for four-dimensional supergravity (expanding with
respect to both the bosonic and fermionic coordinates) and has used it for
computing super-$b_4$ coefficients \cite{superb4}.
Atick and Dhar \cite{atick} have used normal coordinate methods to obtain
an expansion with respect to the fermionic  $\th$ coordinates for the
Green-Schwarz superstring
action in the background of 10-dimensional supergravity. This expansion was
subsequently used for studying  properties of the corresponding $\s$-model
action \cite{sigmamodel} and for
computing $\b$-functions \cite{betafunction}. In the present work we consider
another application.

For many years, one small bone sticking in the collective throat of superspace
practitioners has been the inability to derive by purely superspace methods the
so-called
density formula --- the expression which starting from a superspace
supergravity or matter-supergravity action  of the form
\beq
S= \int  d^d x d^N \th E^{-1} {\cal L}
\eeq
gives the corresponding component action. Here $E = {\rm sdet} E_A^{~M}$ is the
vielbein determinant.
To obtain the component action various techniques have been used on an
ad  hoc basis: a Noether-like
procedure which starts with the knowledge of one term in the component action
and
obtains the rest by (local) supersymmetry transformations \cite{superspace};
by explicit
(covariant) $\th$ expansions
 \cite{bagger}; or some other ingenious acrobatics relying in part on explicit
Wess-Zumino gauge information \cite{buchbinder}. A special case, that of
$(2,2)$ supergravity was recently  treated by purely superspace methods
\cite{measures}, but again relying on some specific properties of the
system.  Recently, Gates \cite{ectoplasm} has proposed a method for obtaining
locally supersymmetric component actions by considerations involving the
topology of superspace.  A variant of that method, as well as a brief account
of the present work, can be found in \cite{ectonorcor}.

The idea of using normal coordinate expansions is simple:  one way to obtain a
component action in terms of the superspace action is to just expand the latter
in powers of the  Grassmann coordinates $\th^\a$, and  simply do the
$\th$-integration. However, in this primitive form, one is led  in general to
noncovariant results, and field redefinitions are needed to recast them in
terms
of (component supergravity) covariant objects. A normal coordinate expansion
with
respect to suitable fermionic variables can obviate this problem because it
leads in
general to covariant expressions.
As mentioned above, McArthur \cite{mcarthur} has given a normal coordinate
expansion of the vielbein. However, he expands $E_A^{~M}(x, \th )$ with respect
to both bosonic and fermionic coordinates around the point $x=0$, $\th =0$ and
therefore his expansion is not suitable for our purpose.
In their work on the expansion of the superstring action \cite{atick},
Atick and Dhar  considered $\th$ expansions leading to covariant expressions
which
were sufficient for their, and
subsequent, purposes, namely  expansions for the quantity $V_i^A \equiv \pa_i
Z^M E_M^{~A}$ as well as various covariant field strengths. It was not
necessary to separately expand the
vielbein itself.  However, their  procedure can be easily adapted for the case
at hand  and leads to a suitable expansion of the vielbein and hence of the
vielbein determinant, and  a purely superspace derivation of the component
density formula.

The outline of our paper is as follows: In section 2 we  present the
modified  expansion of the  inverse vielbein
 ${E_M}^A (x, \th )$  with respect to normal
 coordinates around any point, although in the applications we will generally
expand only with respect to fermionic normal coordinates. Because of the
Grassmann nature of these coordinates, the expansion cuts off after a finite
number of terms.
The subsequent sections apply the general formalism to several specific
examples: two-dimensional $(1,0)$, $(1,1)$ and $(2,0)$ supergravity,
and four-dimensional $N=1$ minimal supergravity. Other cases can
be treated straightforwardly. We observe that as the number of
supersymmetries, and therefore the number of spinor coordinates, increases,
so does the order of the normal coordinate expansion, leading to some
algebraic complexity. Thus, for example, for $N=1$ supergravity in
four-dimensions one would a priori need to expand to fourth (and to some
extent,  fifth) order. We avoid this, in section 7, by expanding in two steps:
 first with
respect to (dotted) spinor variables only (requiring only second order
expansion),  leading to the familiar reduction of the full superspace
integral to a chiral integral, and then again with respect to the
remaining spinor coordinates (requiring again only second order
expansion). This procedure is first illustrated  in section 6 for the simpler
 case of $(2,0)$ supergravity.  Section 8 contains our
conclusions. The constraints for the various supergravity theories
are given  in the Appendix.

\sect{Normal Coordinate Expansion}

Our procedure is straightforward and basically not different from that in ref.
\cite{atick}: we parametrize superspace by coordinates $(x^m , y^\a )$
where the  $y^\a$  describe   a tangent
vector   in a fermionic direction at a point  with coordinates $(x^m,0) $.
(Generically we use
lower-case Greek letters for both dotted and undotted spinor indices,
 wherever no distinction needs to be made.)
More precisely, we choose the origin of our normal coordinate system
to be an arbitrary point with coordinates $z^M = (z^m, z^\mu)$.   We
parametrize a neighbourhood of the origin by normal coordinates $y^A= (y^a,
y^\a)$. The point with coordinates $y^A$ is reached by parallel transport from
the origin. Eventually we identify $z^m = x^m$ and   $y^\a = \th^\a$ and then
 set $z^\mu = y^a = 0$. The normal coordinate expansion that we use gives the
vielbein in Wess-Zumino gauge, as can be seen by examining its components and
their spinor derivatives.   WZ gauge for the supergravity fields is completely
equivalent to this normal coordinate expansion.
The superspace action can then be written in the form
\beq
S= \int d^d x ~d^N y~ E^{-1}(x, y) {\cal L}(x,y)
\eeq
and after expansion in powers of $y^\a$, the fermionic integration is trivial,
using the
basic Grassmann integration rule $\int dy_\a y^\b = \d_\a{}^\b$ or
just picking out the highest-order
terms in $E^{-1}{\cal L}$.  The expansion of $E^{-1}$ automatically gives the
usual factor of $e^{-1} = det\ {e_m}^a$.

More generally, we expand with respect to  some subset of the
coordinates:
We first divide the coordinates into sets $(z^i, z^s)$ and $(y^i, y^s)$.
After using the algorithm we set $z^s=y^i =0$ (in fact  the
$z^s$ can be set equal to some arbitrary constants) and use $(z^i, y^s)$ as a
new coordinate system. The surviving $y$'s
are the normal-gauge-fixed coordinates, while the remaining $z$'s are
still arbitrary coordinates.
  Such gauges are useful, e.g., for
``compactification", where expansions are made in some of the coordinates,
while coordinate invariance is still desired in the remaining coordinates.
The most familiar case is  that of Riemann normal coordinates, where we set all
the $z$'s to vanish, and keep all the $y$'s as our new coordinates.  A
more relevant example is Gaussian normal coordinates, where we
choose $y$ to be a single timelike coordinate, and $z$ the spacelike
coordinates, by setting $z^0=y^i=0$.  This construction then gives the
timelike gauge $g_{m0}=\eta_{m0}$, fixing the time coordinate while
leaving the space coordinates arbitrary.  For  covariant
component expansions in
supersymmetry, the idea is to fix  the fermionic   coordinates, while
maintaining coordinate invariance in the bosonic coordinates.

Superspace integrations are often  performed over
``subsuperspaces" parametrized by the usual spacetime coordinates
plus a subset of fermionic coordinates, for example
 (anti)chiral superspace for $D=4$, $N=1$.   In such
cases normal coordinate expansions can be used (1) to reduce a
  full superspace action, e.g. $ \int d^4 x\ d^4 \theta\ E^{-1}{\cal L}$
 to a subsuperspace action, e.g.  $
\int d^4 x\ d^2 \theta\ {\cal{E}}^{-1}{ {\cal L}_{ch}}$  , and (2) to derive
the
component expansion of the
subsuperspace action e.g. $ \int d^4 x e^{-1}{L}$ .
 In fact, these are the two steps we use  in practice to
evaluate the component expansion of  a full  superspace action.
We  can interpret  the chiral integral as being obtained from  the full
integral
 by expanding  (and integrating
over) only with respect to the $\bar{\theta}^{\dot{ \alpha}}$.
  (Of course, in any
situation coordinates can be integrated out one at a time, but the result will
not
always be simple.  The reduction  produces a manifestly
covariant result only when scalars, e.g. chiral scalars  in curved  $D=4$,
$N=1$ superspace,
can be defined on such subsuperspaces.) The
procedure in both steps is the same; the only difference is the choices
of the various sets of coordinates ($z$'s and $y$'s).  For our chiral
superspace example, which we will discuss in more detail below, the
first step expands with respect to $\bar{\theta}^{\dot{\alpha}}$,
dividing up the superspace coordinates as
$(z^m,z^\mu;y^{\dot{ \alpha}})$, while the second step expands with
respect to $\theta^\alpha$, dividing up the coordinates as
$(z^m,z^{\dot{\mu}};y^\alpha)$.  (A familiar analogue is Euler angles,
coordinates for SU(2) group space defined by a succession of rotations
through each angular coordinate.)  Of course, performing the
steps in succession, rather than all at once,  produces a different
coordinate choice since the covariant derivatives that enter in the parallel
transport  do not commute (even
in flat superspace), so the expression for the component lagrangian may
differ by a total derivative, although the component action is the same.

The point with coordinates $(y^a , y^\a)$ is obtained by an {\em active}
coordinate transformation --- parallel transport --- from the point with
coordinate
$(z^m, z^\mu)$.
 In particular, under such a transformation,
the superspace covariant derivative
\beq
\Del_A = E_A^{~M} D_M +\o _{A} M
\eeq
with $D_M = (\pa_m , D_\m)$ and  $\o_A M = \o_{AB}^{~~~C} M_C^{~B}$,   $M$
a Lorentz generator,
transforms  in standard fashion as
\beq
\Del_A \rightarrow e^{y \cdot \Del} \Del_A e^{- y \cdot  \Del}
\label{eq:transf}
\eeq
with $y \cdot \Del = y^B \Del_B$ and  $y^B = E_M^{~B} y^M$.
Infinitesimally\footnote{ One normally introduces an affine parameter $t$, and
considers expansions in powers of $t$, with the affine parameter being set to
unity, $t=1$, at the end
of the calculation. We dispense with introducing it explicitly.}
 this gives
\beq
\d \Del_A = [ y^B \Del_B , \Del_A] = y^B (T_{BA}^{~~~C} \Del_C +R_{BA} M ) -
(\Del_Ay^B)\Del_B
\eeq
Here $T_{BA}^{~~~C}$ and $R_{BA} $ are the torsion and curvature respectively,
defined by $ \{ \Del_A , \Del_B] = T_{AB}^{~~C} \Del_C +R_{AB}M$ and $R_{AB}M =
R_{ABC}^{~~~~D} M_D^{~C}$.
Under the coordinate transformation tensors undergo the variation
\beq
 \d {\cal T} = y \cdot \Del {\cal T} \label{eq:vartensor}
\eeq
Since by definition of normal coordinates  the tangent vector $y^A$ is parallel
transported, we have
\beq
\d y^A = \d (y \cdot \Del ) =0
\eeq
{}From the infinitesimal variation of the covariant derivative we deduce the
corresponding variations of the vielbein and connection:
\bea
&& E_A^{~M} \d E_M^{~B} = \Del_A y^B - y^C T_{CA}^{~~~B} \nonumber\\
&& E_A^{~M} \d \o_M = y^B R_{BA}           \label{eq:var}
\eea

The finite transformation corresponding to (\ref{eq:transf}) is obtained by
iteration.
For this purpose it is most convenient to consider the variation of the
differential
forms rather than the covariant derivatives. Thus, with the definition
\beq
E^A = dz^M  E_M^{~A} ~~~~,~~~ \o_B^{~A} = dz^M \o_{MB}^{~~~A}
\eeq
 we deduce  from  (\ref{eq:var}),
\beq
 \d E^A = Dy^A +y^C E^B T_{BC}^{~~~A}   \label{eq:varE}
\eeq
where the covariant differential is
\beq
Dy^A \equiv  E^B \Del_B y^A = dy^A-y^B \o_B^{~A} = dy^A -y^B E^C
\o_{CB}^{~~~A} \label{2.11}
\eeq
{}From the variation of the connection we also obtain
\beq
\d Dy^A = -y^B E^C y^D R_{DCB}{}^A   \label{eq:varDy}
\eeq
These equations are, of course, essentially identical to those of ref.
\cite{atick}.

In general, we evaluate the transformed vielbein
\bea
{E'}^A (z; y) &=& E^A +\d E^A + \frac{1}{2!} \d^2 E^A +\frac{1}{3!}\d^3
E^A+......  \non \\
&=& E^B(z) F_B{}^A +(Dy^B)G_B{}^A  \label{FG}
\eea
 where $F$ and $G$ depend explicitly on $T$ and
$R$ and their covariant derivatives, and on $y$.
The procedure is now straightforward;  higher order terms in the expansion are
obtained by iterating the variations in  (\ref{eq:vartensor}, \ref{eq:varE},
\ref{eq:varDy}).  Thus, the second variation is
\bea
\d^2E^A &=& -y^B E^Cy^DR_{DCB}^{~~~~A} + y^C E^B y^D \Del_D T_{BC}^{~~~A}
              \nonumber\\
       &&+ y^C(Dy^B +y^E E^D T_{DE}^{~~~B})T_{BC}^{~~~A}
\eea
and the third is
\bea
\d^3 E^A &=& -y^D (Dy^B+ y^F E^G {T_{GF}}^B) y^C {R_{CBD}}^A
      -y^D E^By^C y^F \Del_F  {R_{CBD}}^A
              \nonumber\\
&&+ y^C (Dy^B + y^F E^G {T_{GF}}^B)y^D \Del_D {T_{BC}}^A +
         y^C E^B y^D y^E \Del_E \Del_D {T_{BC}}^A
        \nonumber\\
&&- y^C y^D E^F y^E {R_{EFD}}^B {T_{BC}}^A +y^C Dy^B y^D \Del_D {T_{BC}}^A \non
\\
&& +y^C y^G(Dy^D + y^E E^F {T_{FE}}^D)
           {T_{DG}}^B {T_{BC}}^A  \non \\
&&+y^C y^F E^D y^E (\Del_E {T_{DF}}^B){T_{BC}}^A
 + y^C y^F E^D {T_{DF}}^B y^E (\Del_E{T_{BC}}^A)
\eea
The fourth variation is:
\bea
\d^4 E^A &=& y^D y^E E^F y^G {R_{GFE}}^B y^C {R_{CBD}}^A - y^D Dy^B y^C y^E
\Del_E  {R_{CBD}}^A
 \non \\
&& - y^D y^F(Dy^E + y^H E^G {T_{GH}}^E){T_{EF}}^B y^C {R_{CBD}}^A \non \\
&&- y^D y^F E^G
y^E
(\Del_E{T_{GF}}^B) y^C {R_{CBD}}^A
 - y^D y^F E^G {T_{GF}}^B y^C y^H \Del_H {R_{CBD}}^A \non\\
  &&  - y^D (Dy^B + y^F E^G {T_{GF}}^B) y^C y^E \Del_E {R_{CBD}}^A  \non \\
&& - y^D E^B y^C y^E y^F \Del_F \Del_E {R_{CBD}}^A \non \\
&& -y^C y^E E^F y^G {R_{GFE}}^B y^D \Del_D  {T_{BC}}^A + y^C Dy^B y^D y^E
\Del_E \Del_D {T_{BC}}^A
  \non \\
&&  + y^C y^F(Dy^E + y^G E^H {T_{HG}}^E){T_{EF}}^B y^D \Del_D {T_{BC}}^A \non
\\
&& +y^C y^F E^H  y^G (\Del_G {T_{HF}}^B) y^D \Del_D {T_{BC}}^A  \non \\
&&  +y^C y^F E^H {T_{HF}}^B y^D y^G \Del_G \Del_D {T_{BC}}^A  \non \\
&&  +y^C(Dy^B + y^F E^G {T_{GF}}^B) y^D y^E \Del_E \Del_D {T_{BC}}^A  \non \\
&& + y^C E^B y^D y^E y^F \Del_F \Del_E \Del_D {T_{BC}}^A \non \\
&& - y^C y^D (Dy^F + y^G E^H {T_{HG}}^F) y^E {R_{EFD}}^B  {T_{BC}}^A \non \\
&& -y^C y^D E^F y^E y^G (\Del_G  {R_{EFD}}^B )  {T_{BC}}^A \non \\
&& -y^C y^D E^F y^E {R_{EFD}}^B y^G \Del_G {T_{BC}}^A \non \\
&& -y^C y^E E^F y^G {R_{GFE}}^B y^D  \Del_D {T_{BC}}^A  \non \\
&& + y^C Dy^B y^D y^E \Del_E \Del_D {T_{BC}}^A \non \\
&&  - y^C y^E y^F E^G y^H {R_{HGF}}^D {T_{DE}}^B {T_{BC}}^A \non \\
&& + y^C y^F Dy^D [y^E (\Del_E {T_{DF}}^B) {T_{BC}}^A  +  {T_{DF}}^B y^E \Del_E
{T_{BC}}^A] \non \\
&& + y^C y^G y^E ( Dy^F + y^I E^H {T_{HI}}^F)  {T_{FE}}^D {T_{DG}}^B {T_{BC}}^A
\non \\
&& +  y^C y^G y^E E^F [ y^H (\Del_H {T_{FE}}^D) {T_{DG}}^B {T_{BC}}^A
    +{T_{FE}}^D y^H (\Del_H {T_{DG}}^B) {T_{BC}}^A \non \\
&& {~~~} +{T_{FE}}^D {T_{DG}}^B y^H (\Del_H {T_{BC}}^A)]
    \non \\
&&  + y^C y^F(Dy^D + y^G E^H {T_{HG}}^D) y^E \Del_E {T_{DF}}^B {T_{BC}}^A \non
\\
&& + y^C y^F E^D y^E y^G(\Del_G \Del_E {T_{DF}}^B ){T_{BC}}^A \non \\
&& + y^C y^F E^D y^E (\Del_E {T_{DF}}^B) y^G \Del_G {T_{BC}}^A  \non \\
&&  + y^C y^F(Dy^D + y^G E^H {T_{HG}}^D) {T_{DF}}^B  y^E \Del_E{T_{BC}}^A \non
\\
&& + y^C y^F E^D y^G (\Del_G {T_{DF}}^B ) y^E \Del_E {T_{BC}}^A \non \\
&& + y^C y^F E^D {T_{DF}}^B y^E y^G \Del_G \Del_E {T_{BC}}^A
\eea
(In the third and fourth variations, we have not collected like terms, for
ease of comparison.)

 For an
action integrated with $d^Ny$ one needs in fact the   variation up to
$\d^{N+1}$  because it  also gives rise to  terms proportional to $y^N$ (times
$Dy$). However, to obtain the
relevant terms in $\d^{N+1} E^A$ for example, one needs only to make
two changes on the right hand side of  $\d^N E^A$ : first, one replaces
 the $E^C$ factors  by $ Dy^C $, and second,
for the terms proportional to $Dy$, one replaces the tensor ${\cal T}$
appearing
in such terms, by $\d {\cal T} = y \cdot \Del {\cal T}$.
There is then no actual need to work out  the variation  beyond
 $\d^N E^A$.

Quite generally, the algorithm can be used for any division
of $z^M$ and $y^A$ into complementary sets. The procedure applies for expansion
about any subset of the coordinates; only the range of the indices changes.
However, to simplify notation we will use indices corresponding
to the special case of expansion of the full superspace over all the
fermionic coordinates, with the understanding that appropriate
modifications can be made for other cases.
Thus, we proceed as follows:
\begin{itemize}
\item
On the right hand side  all quantities  are evaluated  at the
origin
of the normal coordinates, chosen as  $z^M=(z^m,0)$ (and with $dz^\m =0$).
In particular
\bea
E^A(z^m,0) &=& dz^n E_n{}^A(z^m,0) \\
           &=& \left\{
                 \begin{array}{c}
                dz^n e_n{}^a(z^m)  \\
                - dz^n \psi_n{}^\alpha(z^m) \non
                  \end{array}
               \right.
\eea
 Here $e_n{}^a$ is the
component vielbein and
$\psi_n{}^{\alpha} $ the gravitino field.  The minus sign arises because we
usually
define the gravitino from the {\em inverse}, $E_c{}^{\n}| = \psi_c{}^\n$,
 c.f. ref. \cite{superspace}, eq. (5.6.6).

\item
We specialize $y^A$ to be purely spinorial, $(0, y^\a)$ (and also $dy^a =0$).
 We  observe that splitting
up the covariant differential $Dy^\a$ into an ordinary differential $dy^\a$ and
a connection term,  c.f. (\ref{2.11}),  introduces
the latter explicitly as a noncovariant dependence in the normal coordinate
expansion.
However, in any Lorentz invariant quantity,  such as the vielbein determinant,
the dependence drops out. Alternatively,
one can remove it by an additional Lorentz rotation accompanying the transport
of the vielbein
from the origin of the normal coordinates. Keeping it at intermediate stages
provides a check on the calculation, but to simplify the algebra we will simply
drop it.

\item
We use the torsion and curvature constraints appropriate to the supergravity
theory under
consideration, in particular the fact that in $R_{ABC}^{~~~~D}$, $C$ and $D$
must
be
both either spinorial or vectorial.

\item
We write ${E'}^A(z^m,0; 0,y^\a) = E^A(z^m,y^\a) =
 dz^m E_m^{~A}(z, y) + dy^\b E_\b^{~A}(z, y)$ and
read off the normal coordinate expansion of the components from
the coefficients of $dz^m$ and $dy^\b$.
On the right-hand-side of the normal coordinate expansion we have in each term
(aside from those proportional to $Dy^C$)  one factor
 \bea
E^C(z^m,0) &=& dz^m E_m^{~C} = dz^m e_m^{~b} \d_b^C - dz^m \psi_m^{\b}
\d_{\b}^C = e^b [ \d_b^C - \psi_b^{\gamma} \d_{\gamma}^C] \nonumber\\
&\equiv& E^b(z^m,0) \check{E}_b^A(z^m,0)
\eea
with $e^b = dz^m e_m^{~b}$, the component (gravitational) vielbein.
 We thus have
\beq
E^A(z,y) = E^b(z,0)\hat{E}_b{}^A(z,y) +(Dy^\b) \hat{E}_\b{}^A(z,y)
\eeq
where
\bea
\hat{E}_b{}^A &=& \check{E}_b{}^C(z,0)F_C{}^A(z,y)= F_b^A - \psi_b^\g F_\g^A
\nonumber\\
 \hat{E}_\b{}^A&=& G_\b{}^A(z,y)
\eea

Therefore we obtain
\bea
E^{-1} =  {\rm sdet} \pmatrix{
e_m{}^b \hat{E}_b{}^a &  e_m{}^b \hat{E}_b{}^\a\cr
\hat{E}_\b{}^a & \hat{E}_\b{}^\a \cr
}
= e^{-1} \cdot {\rm sdet}   \hat{E}_B^{~A}
\eea
where  $\hat{E}_\b{}^A$ is obtained directly as the coefficient of $Dy^\b$
while
$\hat{E}_b^{~A}$ is obtained by substituting on the right-hand-side of
the normal coordinate expansion $\d_b^c$ and $-{\psi_b}^{\gamma}$,
respectively,
 wherever
${e_m}^c$ and $-{\psi_m}^{\gamma}$  would appear. The expansion involves
now only objects with (component) tangent space indices. We will generally drop
the hat
on $\hat{E}_B^{~A}$. The superdeterminant itself is computed, in standard
fashion:
\bea
{\rm sdet}\pmatrix{
A & B \cr
C & D \cr
}
= \frac{ \det (A-BD^{-1}C)}{\det D}
\eea
although sometimes it is simpler to evaluate it from ${\rm sdet}
 E ={\rm exp} ( {\rm str ~ln}E)$.

\end{itemize}

 In the next section we consider, as a simple illustration of the procedure,
the
density formula for the two-dimensional   $(1,0)$   system.

\sect{Two-dimensional $(1,0)$ Supergravity }

 The $(1,0)$ supergravity case is the simplest to consider. The superspace is
parametrized by two bosonic coordinates $x^m$ and one fermionic
coordinate $y^+$. We consider an action of the form
\beq
S= \int d^2x dy_+ E^{-1}(x,y) {\cal L^+}(x,y)       \label{eq:10action}
\eeq
where $\int dy_+ y^+ = 1$.
 We need the expansion of the vielbein and the
lagrangian to first order in $y^+$.
 For the latter we simply use
\beq
{\cal L^+}(x,y) = {\cal L^+}| +y^+( \Del_+ {\cal L^+}| )
\eeq
where the vertical bar $|$ indicates as usual evaluation at $y^+$=0.
In the  expansion of the vielbein determinant,  for terms
linear
in $y^+$ we need to go up to the second variation of $E^A$ (but keeping only
one term, the one proportional to $Dy^+$, from that variation). We have
\beq
E^A(x,y) = E^A(x,0) +  Dy^A +y^+ E^C T_{C+}^{~~~A}  +
 \frac{1}{2} y^+ Dy^+T_{++}^{~~~A}
\eeq
The supergravity constraints for the torsion that enter here are
\cite{brooks,sugraphity}
$T_{++}^{~~~A} = 2i \d_\pp^{~A}$, $ T_{C+}^{~~~A}  = 2i \d_C^{~+}
\d_\pp^{~A}$. (The complete set of constraints can be found in the Appendix.)
 Using  $E_m^{~+}| =- \psi_m^{~+}$
 we find then
\bea
&&E_m^{~a} = (e_m^{~a} - 2i y^+ \psi_m^{~+})\d_\pp^{~a} = e_m^{~b}(\d_b^a - 2i
y^+\psi_b^+ \d_{\pp}^a)\nonumber\\
&&E_m^{~+} = -\psi_m^{~+}  =- e_m^{~b}\psi_b^+\nonumber\\
&&E_\b^{~a}=   -iy^+ \d_\b^+ \d_\pp^a      \nonumber\\
&&E_\b^{~+}= \d_\b^{~+}
\eea
 We obtain
\beq
E^{-1}  = e^{-1}[ 1-i y^+ \psi_\pp^{~+}]
\eeq
Therefore, the action in (\ref{eq:10action}) is
\bea
S&=& \int d^2x dy_+ e^{-1}[ 1-i y^+ \psi_\pp^{~+}]\cdot
 ( {\cal L^+}| +y^+ \Del_+      {\cal L^+}| )\nonumber\\
&=& \int d^2x~ e^{-1}[  \Del_+ -i  \psi_\pp^{~+}]  {\cal L^+}|
\eea
which is the familiar density formula for this case \cite{brooks}.

\sect{ $(1,1)$ Supergravity}

We turn now to the case of $(1,1)$ supergravity where superspace is
parametrized by two spinorial coordinates, $y^+$ and $y^-$, in addition to the
bosonic ones.
We consider an action of the form
\beq
S= \int d^2x dy_+ dy_- E^{-1}(x,y) {\cal L}(x,y)       \label{eq:11action}
\eeq
We need the normal coordinate expansion to second order in $y$.
The Lagrangian expanded to that order is
\beq
{\cal L}(x,y) = (1 + y^+ \Del_+ + y^- \Del_- + \shalf y^- y^+ [\Del_+, \Del_-])
{\cal L}|
\eeq
 The relevant
constraints are \cite{martin, jim, nishino}
\bea
{T_{++}}^\pp = 2i  &,& {T_{--}}^\mm = 2i\non \\
{T_{+\mm}}^- = \sihalf R &,& {T_{-\pp}}^+ = -\sihalf R
\eea
We find for the expansion of the vielbein (actually, as discussed in
 section 2, $\hat{E}_B^{~A}$)
\bea
&&E_b^{~\pp} = \d_b^{~\pp}-2i y^+ \psi_b^{~+}  -\shalf y^+ y^- \d_b^{~\pp}
R\nonumber\\
&&E_b^{~\mm} = \d_b^{~\mm}-2i y^- \psi_b^{~-}  -\shalf y^+ y^- \d_b^{~\mm}
R\nonumber\\
&&E_b^{~+} =- {\psi_b}^+
        + \sihalf y^-  \d_b^{~\pp} R -\sihalf \d_b^{~\pp} y^+ y^- \Del_+
           R - \squart y^+ y^- \psi_b^{~+} R \nonumber\\
&&E_b^{~-} =-{\psi_b}^- -
            \sihalf y^+ \d_b^{~\mm} R -\sihalf \d_b^{~\mm} y^+ y^- \Del_-
           R - \squart y^+ y^- \psi_b^{~-} R  \nonumber\\
&&E_\b^{~\pp}=-i y^+ \d_\b^\pp \non \\
&&E_\b^{~\mm}=-i y^- \d_\b^\mm \non \\
&&E_\a^{~+} = \squart y^+ y^- \d_\a^+R \non\\
&&E_\a^{~-} = \squart y^+ y^-\d_\a^- R
\eea
with $R$ and $\Del_\a R$ evaluated at $(x,0)$.

{}From this we compute the superdeterminant and find
\beq
E^{-1} = e^{-1}[ 1+i y^+ \psi_\pp^{~+}  + i y^- \psi_\mm^{~-} -\shalf y^+ y^- R
+ y^+ y^-
     ( \psi_\pp^{~+} \psi_\mm^{~-} - \psi_\mm^{~+} \psi_\pp^{~-}) ]
\eeq
The action \reff{eq:11action} is therefore
\bea
S&=& \int d^2x dy_+dy_- e^{-1}[ 1+i y^+ \psi_\pp^{~+}  + i y^- \psi_\mm^{~-}
           -\shalf y^+ y^- R + y^+ y^-
     ( \psi_\pp^{~+} \psi_\mm^{~-} - \psi_\mm^{~+} \psi_\pp^{~-}) ]\non \\
 &&~~~~ ~~~~~~~~~~~~~~~~~~~~~~~~~~  \cdot  (1 + y^+ \Del_+ + y^-
     \Del_- - y^+ y^- \Del_+ \Del_-) {\cal L}| \nonumber\\
&=& \int d^2x~ e^{-1}[  \Del_+ \Del_-  +i \psi_\pp^{~+}\Del_-  +i
\psi_\mm^{-}\Del_+
    +\shalf R - (\psi_\pp^{~+} \psi_\mm^{~-} - \psi_\mm^{~+} \psi_\pp^{~-}) ]
{\cal L}|
\eea
which agrees with the result in \cite{jim}.

\sect{ $(2,0)$  Supergravity}

We move on to the case of (2,0) supergravity, with superspace  parametrized by
 $x^m$, as well as $y^+$, $y^{\dot{+}}$. The action for this case is
\beq
S = \int d^2x dy_+ dy_{\dot{+}} E^{-1}(x,y) {\cal L}^{+ \dot{+}}(x,y)
\label{eq:20action}
\eeq
Again, the Lagrangian is expanded to second order as
\beq
{\cal L}^{+ \dot{+}}(x,y) = (1 + y^+ \Del_+ + y^{\dot{+}} \Del_{\dot{+}} +
\shalf  y^{\dot{+}}y^+
       [\Del_+, \Del_{\dot{+}}]) {\cal L}^{+ \dot{+}}|
\eeq
The required torsions, from  the appendix, are
\bea
{T_{+ \dot{+}}}^\pp =2i && \non \\
{T_{\pp \mm}}^+ =- \S^+  &,& {T_{\pp \mm}}^{\dot{+}} =- \S^{\dot{+}} \non \\
{T_{+ \mm}}^+ = iG_\mm   &,& {T_{\dot{+} \mm}}^{\dot{+}} = -iG_\mm \non
\eea
The vielbein components (for $\hat{E}$)  are:
\bea
{E_\pp}^\pp &=& 1 - 2i (y^+ {\psi_{\pp}}^{\dot{+}} + y^{\dot{+}}
{\psi_{\pp}}^+) \non \\
{E_\mm}^\pp &=&  - 2i (y^+ {\psi_{\mm}}^{\dot{+}} + y^{\dot{+}} {\psi_{\mm}}^+)
               - 2   y^+ y^{\dot{+}} G_\mm  \non \\
{E_\pp}^\mm &=& 0 \non \\
{E_\mm}^\mm &=&1 \non \\
{E_+}^\pp &=& -iy^{\dot{+}} \non \\
{E_{\dot{+}}}^\pp &=& -iy^+ \non \\
{E_+}^\mm &=&0 \non \\
 {E_{\dot{+}}}^\mm &=&0 \non \\
{E_\pp}^+ &=& -{\psi_\pp}^+ \non \\
{E_\pp}^{\dot{+}} &=& -{\psi_\pp}^{\dot{+}} \non \\
{E_\mm}^+ &=& -{\psi_\mm}^+ - i y^+ G_\mm  + i y^+ y^{\dot{+}}
\S^+ \non \\
{E_\mm}^{\dot{+}} &=& -{\psi_\mm}^{\dot{+}}+ i y^{\dot{+}} G_\mm
- i y^+ y^{\dot{+}}
  \S^{\dot{+}} \non \\
{E_\b}^\a &=& \d_\b^\a
\eea
  We find for the superdeterminant
\beq
E^{-1} = e^{-1}[ 1 -i y^+ {\psi_\pp}^{\dot{+}}  - i y^{\dot{+}} {\psi_\pp}^{+}]
\eeq
and therefore the action \reff{eq:20action} becomes
\bea
S&=& \int d^2x dy_+dy_{\dot{+}} e^{-1}[ 1-i y^{\dot{+}} \psi_\pp^{~+}  - i y^+
\psi_\mm^{~\dot{+}} ]  \non \\
 &&~~~~~~~~~~~~~~~~~~~~~~
\cdot      (1 + y^+ \Del_+ + y^{\dot{+}} \Del_{\dot{+}} + \shalf
y^{\dot{+}}y^+
       [\Del_+, \Del_{\dot{+}}] ){\cal L}^{+ \dot{+}}|  \non \\
&=& \int d^2x~ e^{-1}\{ \shalf [\Del_+, \Del_{\dot{+}}]  + i
\psi_\pp^{~+}\Del_+ -
            i \psi_\pp^{~\dot{+}}\Del_{\dot{+}}\}{\cal L}^{+ \dot{+}}|
\label{s20}
\eea
which is consistent with the density formula in \cite{brooks}.

We note for future reference, using $ \shalf  [\Del_+, \Del_{\dot{+}}] =\Del_+
\Del_{\dot{+}}
-i \Del_\pp$
and evaluating  at the origin of normal coordinates with $\Del_\pp = {\bf
D}_\pp
+ \psi_\pp^+ \Del_+ +\psi_\pp^{\dot{+}}\Del_{\dot{+}}$ ( the ${\bf D_\pp}$
term gives a total derivative that we drop), that the action can be rewritten
as
\beq
S=  \int d^2x dy_+dy_{\dot{+}} e^{-1} [ \Del_+ -2i \psi_\pp^{\dot{+}}]
\Del_{\dot{+}}
{\cal L}^{+ \dot{+}}|
 \eeq
suggesting that the density formula could be obtained in a two-step process
going through the intermediate stage of a ``chiral'' integral. We discuss this
in the next section.

\sect{A Chiral Decomposition in $(2,0)$ Supergravity}
In this section, we recover the  last density formula in the previous section
by
going through an intermediate stage of rewriting the $(2,0)$ superspace
integral
as a $dy_+$ integral of a ``chiral'' integrand.

As discussed in section 2, instead of considering
 a normal coordinate expansion around
 the origin  $(z^m, 0, 0)$ in a direction parametrized by $y^+$ and
$y^\pd$, we  can consider  an expansion in a single fermionic
coordinate, with the origin $(z^m, z^+, 0)$, in
the direction parametrized by $y^\pd$.
The expansion of the $(2,0)$ vielbein is similar to that in the $(1,0)$ theory:
\beq
E^A(z, y^\pd) =  E^A +  Dy^\pd \d_\pd^{~A} +y^\pd E^C
T_{C \pd}^{~~~A}  +
 \frac{1}{2} y^\pd Dy^\pd T_{\pd \pd}^{~~~A}
\eeq
with the quantities  on the right hand side evaluated at $y^\pd =0$. Note that
 in $(2,0)$ theory  $T_{\pd \pd}^{~~~A}=0$.

The analogue of (2.17) is $E^C = dz^{\rm m}[{E}_{\rm m}{}^{\rm b}\d_{\rm b}^C +
E_{\rm m}{}^\pd \d_\pd^C ]$  where we have introduced roman letters to
indicate collectively space-time and $+$ spinor indices:
 ${\rm m}=(m, +)$, ${\rm b} = (b, +)$.
We use a modification of the procedure explained in (2.17) and (2.18) by
extracting from the full vielbein matrix a factor
\bea
{E}_{\rm m}{}^{\rm a} =
\pmatrix{
E_m{}^a & E_m{}^\a \cr
E_\m{}^a & E_\m{}^\a \cr
}
\eea

The entries in ${E}_{\rm m}{}^{\rm a}$
are the vielbein components evaluated at  $z^\pd =0$. We have then
(ellipses indicate irrelevant terms)
\bea
E^{-1} = {\rm sdet}
\pmatrix{
{E}_{\rm m}{}^{\rm a}+y^\pd  {E}_{\rm m}{}^{\rm c}T_{{\rm c}\pd}^
{~~~{\rm a}}  ~~~&~~~ \cdots \cr
0~~~ &~~~ 1 \cr
}\eea
which equals
\beq
{\rm sdet} {E}_{\rm m}^{~\rm c}  \times {\rm sdet}
[\d_{\rm c} ^{~\rm a} +y^\pd T_{{\rm c} \pd}{}^{\rm a} ]
\eeq
But the second superdeterminant is unity,  c.f.  (A.3), and we simply obtain
$\tilde{E}^{-1} = {\rm sdet} {E}_{\rm m}^{~\rm c} $.
Consequently we have the reduction to a chiral density formula
\beq
\int d^2x dy_+ dy_\pd E^{-1} {\cal L^{+ \pd}} =
\int d^2x dy_+\tilde{ E}^{-1} \Del_\pd  {\cal L^{+ \pd}}  | _{y^\pd =0}
\eeq

We proceed now to the next step in the reduction, using an expansion with
respect to $y^+$ similar to the one above
\beq
{E}^{\rm a}(x,y^+) = {E}^{\rm a}  +  Dy^+ \d_+^{~\rm a} +
y^+ {E}^{\rm c} T_{{\rm c} +}{}^{\rm a}  +
 \frac{1}{2} y^+ Dy^+ T_{++}{}^{\rm a}
\eeq
We find
\bea
\tilde{E}^{-1} &=& e^{-1} {\rm sdet} \pmatrix{
1-2i y^+{\psi}_\pp^\pd & -2iy^+  {\psi}_\pp^\pd \cr
0 & 1 \cr
} \nonumber\\
&=& e^{-1}(1-2i y^+ {\psi}_\pp^\pd)
\eea
When this is substituted in the chiral integral,
we recover the final expression in the previous section.

\sect{Four-dimensional $N=1$ Supergravity}

As a final example we present the derivation of the density formula
for this case. The same derivation will work for the density formula of
 $(2,2)$ supergravity in two dimensions \cite{measures}. We will consider the
case of minimal, $n= - 1/3$ supergravity, but the same method should
work for any value of $n$. $N=1$ superspace is parametrized by
 $(x^m, \th^\m,
\bar{\th}^{\dot{\m}}) $ and corresponding covariant derivatives. The complete
 set of  constraints defining the torsions and curvatures, is given in the
Appendix.

We will use the same approach as in section 6
for $(2,0)$ supergravity, namely reduce first the $d^4 \th$ integral to
a $d^2 {\th}$  chiral integral, and then reduce the latter.
 In this manner, it is
not necessary to use the normal coordinate expansion to fourth order, as
a direct reduction would require, but only to second order.
 We consider
therefore an expression of the form
\beq
S=\int d^4x d^2\bar{\th}d^2\th E^{-1} {\cal L}= \int d^4x d^2 \th
\int d^2 \bar{\th}E^{-1} {\cal L}
\label{7.1}
\eeq
and expand the integrand  with respect to normal coordinates $y^\ad$ around
an origin with coordinates $(z^m,  {z}^\m ,0)$.
We split up the vielbein components $E_M^{~A}$ into components
$E_{\rm m}^{~{\rm a}}$, $E_{\rm m}^{~\ad}$, $E_{ \dot{\m}}^{~{\rm a}}$ and
$E_{\dot{\m}}^{~\ad}$ where we use roman letters to indicate collectively
space-time indices as well as undotted spinor indices, e.g. ${\rm a} = (a,
\a)$.

We need the
expansion to second order in $y^\ad$, and the
 following nonzero torsions and curvatures  (see
Appendix A)
\bea
T_{\a \ad}^{~~b} &=&i \d_\a^{~\b}  \d_\ad^{~\bd} ~~~~~~,
{}~~~b\equiv (\b \bd) \\
T_{\a b}^{~~\g}&=&i C_{\a \b} G^\g_{~\bd} ~~~~,~~~
T_{\a b}^{~~\dot{\g}}= -iC_{\a \b} \d_\bd^{~\dot{\g}} R \non\\
R_{\d \a \g}^{~~~\a}&=& -2 \bar{R}(M_{\d \a})_\g^{~\a}
\label{7.4}
\eea
where the Lorentz generator is
\beq
(M_\d ^\a)_\g^{~\b}=\d_\d^{~\b} \d_\g^{~\a} -\frac{1}{2}\d_\d^{~\a} \d_\g^{~\b}
\label{7.5}
\eeq
 (Torsions with an odd
number of spinor indices vanish, as do, here,  the curvatures  with a vector or
dotted spinor last index.)

 Once more we modify the
discussion  in section 2,   extracting a factor
 ${E}_{\rm m}^{~{\rm b}}(z^m,  {z}^{{\m}} ,0)$
with ${\rm m}= (m
, {\m})$ and
 ${\rm b}= (b, \b)$.  (One has to pay some attention to  minus signs
when bringing
ahead of spinors some  spinorial components  of   ${E}_{\rm m}^{~{\rm
b}}$).
We compute
\beq
E^{-1}(z^m, z^{{\m}},0; 0,0,{y}^\ad ) =  \tilde E^{-1} {\hat{E}}^{-1}
\label{E}
\eeq
 where
\bea
\tilde E^{-1}  =  sdet \pmatrix{  E_m{}^a & E_m{}^\alpha   \cr
        E_\mu{}^a & E_\mu{}^\alpha  \cr}  \label{det}
\eea
and $ \hat{E}^{-1} = sdet \hat{E}_B{}^A$, with
\bea
 \hat E_B{}^A = \pmatrix{
        \d_b^a -i \bar y^\ad \psi_b^\a -\frac{1}{2} \d_b^a \bar y^2 R
                & -\psi_b^\a +i \bar y^\gd C_{\gd\bd}  \d_\b^\a R+\cdots
                & - \psi_b^\ad +\cdots \cr
        -i\bar y^\ad \d_\b^\a  +\cdots
                & \d_\b^\a +\bar y^2 \d_\b^\a R
                & \cdots \cr
        0 & 0 & \d_\bd^\ad + \frac{1}{2}\d_\bd^\ad \bar y^2 R \cr}
\label{7.6}
\eea
where the ellipses indicate irrelevant terms.

We find, evaluating the superdeterminant
\beq
E^{-1} = \tilde{E}^{-1} [1-\bar{y}^2
{R} ] \label{7.7}
\eeq
  where $\bar y^2 = \frac{1}{2}\bar{y}^{\dot{\alpha}} \bar{y}_{\dot{\alpha}}$.

At the same time, the expansion of the lagrangian ${\cal L}$ is
\beq
{\cal L}= [ 1+ \bar{y}^\gd \bar{\Del}_\gd +\frac{1}{2} \bar{y}^\gd \bar{
y}^{\dot{\d}}
\bar{ \Del}_{\dot{\d}} \bar{\Del}_{\gd}  ] {\cal L}
|_{\bar{\th} =0} \label{7.8}
\eeq
and the action becomes
\beq
S= \int d^4x d^4 \th E^{-1} {\cal L} =\int d^4x d^2 {\th}
\tilde{E}^{-1}
( \bar{\Del}^2+{R} ){\cal L}| ~.\label{7.9}
\eeq
$ \tilde E^{-1}$ is the measure of the chiral subspace.  (One can
 verify in the $n=-1/3$ theory  that $\tilde{E}^{-1} ={\phi}^3$
 where $\phi$ is the superconformal  compensator.)

The same procedure can be applied to the next step.  The coordinate
restriction is now the antichiral one
\beq
 z^\mu \ =\ 0 \ ,\ (y^a, y^{\dot{\alpha}}) \ =\  0
\eeq
 We first need to evaluate $ E_{\rm m}{}^{\rm a}$ , now treated as a function
of the new $z^M$ and $y^A$.  We can use the result of the previous
calculation by: (1) replacing (\ref{7.6}) with the hermitian conjugate
(effectively just switching dotted and undotted indices), and (2)
deleting the second row and column before taking the
superdeterminant, so we get the contribution to the smaller
superdeterminant of (\ref{det}). The result is
\bea
 \tilde E^{-1}  =  e^{-1} [ 1 - iy^\a \bar{\psi}_{\a \ad}{}^\ad -
y^2 (  3 \bar{ R} + \frac{1}{2} C^{\a \b} \bar{\psi}_{ \a \ad}
{}^{(\ad
} \bar{\psi}{}_{ \b \bd}{}^{\bd )}
 ) ]
\eea
 Comparing to (\ref{E}) we identify   the new $\tilde
E^{-1}$  as $e^{-1}=det\ e_m{}^a$  and the rest as the new $\hat
E^{-1}$.  The expansion of the Lagrangian is similar to the previous case
(just switching
$\bar y\to y$), giving the final result
\beq
 S \ =\  \int d^4 x\ d^4 \theta\ E^{-1}{\cal L}
        \ =\  \int d^4 x\ {\rm e}^{-1}{\cal D}^2 ({\Bar \de}{}^2 \,+\, R )
{\cal L}|
        \ =\   \int d^4 x\ {\rm e}^{-1}{\cal D}^4 {\cal L}|
\eeq
 where we have defined a superdifferential operator, the ``chiral
density projector" ${\cal D}^2$, which (for the present case of  old
minimal supergravity) takes the form
\beq
 {\cal D}^2 \ \equiv \  \Del^2 \,+\, i  {\bar \psi}{}^{ \a \ad}{}_{\dot
{\a}}
\Del_{\a} \, +\, 3 \,{\bar R} \, +\, \frac{1}{2} C^{\a \b} {\bar \psi}_{\a \ad}
{}^{( \dot {\a}} \, {\bar \psi}{}_{\b \bd}{}^ {\bd )}
\eeq
 and the general density projector
${\cal D}^4\equiv{\cal D}^2(\Bar\Del {}^2+R)$.  We could use  instead the
complex conjugate $\bar{\cal D}^4=\bar{\cal D}^2(\de^2+\bar R)$,
which differs only by a total space-time derivative. The final result is the
standard density formula for minimal four-dimensional  $N=1$
supergravity.

\sect{Conclusions}

The normal coordinate procedure we have used in this paper for
deriving the density formula is
straightforward and can be applied to any supergravity theory.
Although straightforward, it is not necessarily the
most economical way for obtaining the component  supergravity-matter
action; one has to go through the two stages of first computing the
vielbein, and then the superdeterminant, and algebraically this can be
quite complicated. Obviously, in any specific case, or  even in general,
some short-cut or particular trick may be preferable. However, we feel
that in principle one should have an unambiguous, and extra-assumption-free
procedure for deriving the result, relying on nothing more than the general
properties of superspace. We believe the normal coordinate expansion
provides the appropriate technique.

We conclude with the following remark. In ref. \cite{superspace}, section
5.6.b,
a  closely related technique for obtaining  $\th=0$ components
of covariant derivative was developed. It also allows one, in principle,
to extract from these components an expansion for the vielbein. However,
we believe that the normal coordinate expansion we have used in this paper
 is more systematic and simpler.

\appendix
\section {{\bf Appendix: Supergravity Constraints}}

\setcounter{equation}{0}

  We list here our conventions and the constraints for the  supergravity
theories discussed in the main text.
In general we follow the conventions used in \cite{superspace}.
The constraints for the various cases are listed below.

{\bf (1,0) Supergravity:}
\bea
\{\Del_+ , \Del_+ \} = 2i \Del_\pp &,&  [\Del_+, \Del_\mm ] = -2i \S^+ M \non
\\
\{\Del_+ , \Del_\pp\} = 0 &,&  [\Del_\pp, \Del_\mm ] = - \S^+ \Del_+ + R M
\eea
where $R= 2 \Del_+ \S^+$.

{ \bf (1,1) Supergravity:}
\bea
\{\Del_+, \Del_+\} &=& 2i \Del_\pp  ~~~,~~~~
 \{\Del_-, \Del_- \} = 2i \Del_\mm  \non \\
\{ \Del_+, \Del_- \} &=& R M   \\
{[}\Del_+,\Del_\mm ] &=& i (\Del_-R)M +\sihalf  R \Del_-  ~~~,~~~~
{[}\Del_-, \Del_\pp ] = i (\Del_+R)M -\sihalf R \Del_+   \non
\eea

{\bf (2,0) Supergravity:}
\bea
\{\Del_+, \Del_+ \} &=& 0 ~~~,~~~~
 \{\Del_{\dot{+}}, \Del_{\dot{+}} \} = 0\non \\
\{\Del_+, \Del_{\dot{+}} \} &=& 2i \Del_\pp  \\
{[}\Del_+, \Del_\mm] &=& i G_\mm \Del_+ - 2i(\Del_+ G_\mm )M \non \\
{[}\Del_{\dot{+}}, \Del_\mm] &=& - i G_\mm \Del_{\dot{+}}
+2i ( \Del_{\dot{+}} G_\mm )M \non \\
{[}\Del_\pp, \Del_\mm] &=& ( \Del_{\dot{+}} G_\mm ) \Del_+
       - (\Del_+ G_\mm )\Del_{\pd}+([\Del_+ , \Del_{\dot{+}} ]G_\mm )M  \non
\eea

{\bf (2,2) Supergravity:}
\bea
\{\Del_+, \Del_{\dot{+}} \} &=& 2i \Del_\pp ~~~,~~~~
 \{\Del_-, \Del_{\dot{-}} \} = 2i
\Del_\mm \non \\
\{\Del_+, \Del_-\} &=& - \bar{R} \bar{M} ~~~,~~~~ \{\Del_{\dot{+}},
\Del_{\dot{-}} \}
= RM \non \\
{[}\Del_+, \Del_\mm] &=& -\sihalf \bar{R}
\Del_{\dot{-}}   -i (\Del_{\dot{-}} \bar{R}) \bar{M} \non \\
{[}\Del_{\dot{+}}, \Del_\mm] &=&\sihalf R \Del_-  + i (\Del_-R)M \non \\
{[}\Del_-, \Del_\pp] &=&  \sihalf \bar{R}
\Del_{\dot{+}}  -i (\Del_{\dot{+}} \bar{R}) \bar{M}\non \\
{[}\Del_{\dot{-}}, \Del_\pp] &=&  -\sihalf R \Del_+ +i (\Del_+R)M
\eea

{\bf N=1 Four-Dimensional Supergravity:}
\bea
\{ \Del_\a, \Del_\b \} &=& -2 \bar{R}M_{\a \b} ~~~~,~~~
\{\Del_\a , \Delb_\ad \} = i \Del_{\a \ad} \\
{[} \Del_\a , \Del_{\b \bd}{]} &=& -iC_{\a \b}[ \bar{R} \Delb_\bd -G^{\g}_\bd
\Del_\g ]
\non \\
&&+iC_{\a \b}[ \bar{W}_{\bd \dot{\g}}^{~~\dot{\d}}
\bar{M}_{\dot{\d}}^{~\dot{\g}} -
(\Del^\g G_{\d \bd})M_\g^{~\d} ] -i (\Delb_\bd \bar{R})M_{\a \b} \non \\
{[}\Del_{\a \ad}, \Del_{\b \bd}] &=& [ C_{\ad \bd}W_{\a \b}^{~~\g} +
C_{\a \b}(\Delb_\ad G^\g_{~\bd}) - C_{\ad \bd}(\Del_\a R) \d_\b^{~\g}]\Del_\g
 +C_{\a \b} G^\g_{~\bd}\Del_{\g \ad} \non\\
&&+[C_{\ad \bd}(\Del_\a W_{\b \g}^{~~\d} +(\Delb^2 \bar{R}
+2R\bar{R})C_{\g \b}\d_\a^{~\d}) -C_{\a \b}(\Delb_\ad \Del^\d G_{\g \bd})]
M_\d^{~\g} +h.c. \non
\eea

\newpage

\end{document}